\RequirePackage{fix-cm}
\documentclass[pdflatex,sn-mathphys]{sn-jnl}

\usepackage[T1]{fontenc}
\usepackage{lmodern}
\usepackage{fixmath}
\usepackage{amsmath}
\usepackage{mathrsfs}
\usepackage{microtype}
\usepackage{xurl}
\theoremstyle{thmstyleone}

\theoremstyle{thmstyletwo}

\theoremstyle{thmstylethree}

\begin{document}

\title[Spin-Orbital Hall Nano-Oscillators using PtCr/NiFe]
{Spin-Orbital Hall Nano-Oscillators using PtCr/NiFe}

\author*[1]{\fnm{Utkarsh} \sur{Shashank}}
\author[1,5,6]{\fnm{Akash} \sur{Kumar}}
\author[7,8]{\fnm{Daegeun} \sur{Jo}}
\author[2]{\fnm{Thi Ngoc Anh} \sur{Nguyen}}
\author[1]{\fnm{Jong-Guk} \sur{Choi}}
\author[1]{\fnm{Sambit} \sur{Ghosh}}
\author[3]{\fnm{Micha\l{}} \sur{Strach}}
\author[4]{\fnm{Lunjie} \sur{Zeng}}
\author[4]{\fnm{Andrew B.} \sur{Yankovich}}
\author[1]{\fnm{Roman} \sur{Khymyn}}
\author[1]{\fnm{Ahmad A.} \sur{Awad}}
\author[4]{\fnm{Eva} \sur{Olsson}}
\author[7,8]{\fnm{Peter M.} \sur{Oppeneer}}
\author*[1,5,6]{\fnm{Johan} \sur{\AA kerman}}
\email{utkarsh.shashank@physics.gu.se,johan.akerman@physics.gu.se}

\affil[1]{\orgdiv{Applied Spintronics Group, Department of Physics},
\orgname{University of Gothenburg},
\orgaddress{\city{Gothenburg}, \postcode{412 96}, \country{Sweden}}}

\affil[2]{\orgname{Institute of Materials Science, Vietnam Academy of Science and Technology},
\orgaddress{18 Hoang Quoc Viet, Nghia Do, \city{Hanoi}, \country{Vietnam}}}

\affil[3]{\orgdiv{Chalmers Materials Analysis Laboratory},
\orgname{Chalmers University of Technology},
\orgaddress{\city{Gothenburg}, \postcode{412 96}, \country{Sweden}}}

\affil[4]{\orgdiv{Department of Physics and Astronomy},
\orgname{Chalmers University of Technology},
\orgaddress{\city{Gothenburg}, \postcode{412 96}, \country{Sweden}}}

\affil[5]{\orgdiv{Research Institute of Electrical Communication},
\orgname{Tohoku University},
\orgaddress{\street{2-1-1 Katahira, Aoba-ku},
\city{Sendai},
\postcode{980-8577},
\country{Japan}}}

\affil[6]{\orgdiv{Center for Science and Innovation in Spintronics},
\orgname{Tohoku University},
\orgaddress{\street{2-1-1 Katahira, Aoba-ku},
\city{Sendai},
\postcode{980-8577},
\country{Japan}}}

\affil[7]{\orgdiv{Department of Physics and Astronomy},
\orgname{Uppsala University},
\orgaddress{\city{Uppsala}, \country{Sweden}}}

\affil[8]{\orgname{Wallenberg Initiative Materials Science for Sustainability (WISE)},
\orgaddress{\city{Uppsala}, \country{Sweden}}}
\abstract{

The orbital Hall effect provides a promising route for generating angular-momentum currents beyond conventional spin Hall physics. PtCr alloys exhibit unusually large current-induced torques, but the contribution of orbital transport and the ability of these torques to sustain coherent nonlinear magnetization dynamics remain unresolved. Here we demonstrate spin-orbital Hall nano-oscillators by exploiting a homogeneous heavy-metal/light-metal alloy in which orbital Hall currents generated by Cr are converted by Pt into spin currents, producing giant spin--orbit torques. Using PtCr/NiFe heterostructures, the effective torque efficiency increases from $\sim$0.14 in Pt/NiFe to $\sim$0.40 in Pt$_{0.38}$Cr$_{0.62}$/NiFe despite substantial Pt dilution, enabling coherent auto-oscillations with the threshold current density reduced from $\sim1.07\times10^{12}$ to $\sim4.4\times10^{11}\,\mathrm{A\,m^{-2}}$. First-principles calculations show that Cr alloying suppresses the intrinsic spin Hall conductivity while enhancing the orbital Hall conductivity, and reproduce the observed torque enhancement only when orbital transport is included. 
Our combined experimental and first-principles results show that alloy engineering enables giant spin--orbit torques through an intrinsic orbital-mediated contribution, enabling coherent auto-oscillations without engineered multilayers and establishing a scalable materials platform for low-power nonlinear spintronic and orbitronic devices.
}

\maketitle
\section*{Introduction}\label{sec1}

Electrons carry both spin and orbital angular momentum. Over the past three decades, the generation, manipulation and detection of spin currents via the spin Hall effect (SHE) \cite{Hirsch1999} have established modern spintronics \cite{dieny2020opportunities} as a major research field, enabling magnetization switching \cite{liu2012science}, spin-wave control \cite{fulara2019spin,kumar2025spin}, spin Hall nano-oscillators (SHNOs) \cite{demidov2014nanoconstriction,Awad2017}, and the manipulation of domain walls \cite{emori2013current} and skyrmions \cite{bernstein2025spin}. While these advances have relied predominantly on SHE in heavy 5$d$ transition metals \cite{liu2011spin} with strong spin-orbit coupling (SOC), recent studies have established that orbital angular momentum provides an alternative route for efficient angular momentum transport, extending spin-orbitronics beyond conventional SHE and enabling the use of lighter, earth-abundant 3$d$ transition metals \cite{jo2024spintronics}.

The underlying mechanism is the orbital Hall effect (OHE), first proposed by Bernevig \textit{et al.} \cite{bernevig2005orbitronics} and later predicted in transition metals by Kontani \textit{et al.}~\cite{kontani2009giant}. Subsequent developments based on the concept of orbital texture~\cite{go2018intrinsic} and first-principles calculations established the OHE as a universal phenomenon across 3$d$, 4$d$, and 5$d$ metallic systems~\cite{salemi2022first}. In the OHE, a charge current generates a transverse orbital current carrying orbital angular momentum. Through SOC and exchange coupling, this orbital angular momentum is transferred to the magnetization of an adjacent ferromagnet, either directly or through orbital-to-spin conversion, giving rise to efficient current-induced torques \cite{go2020orbital}. Unlike the SHE, the OHE does not require strong SOC and is predicted to be substantial in a broad range of materials, particularly lighter 3$d$ transition metals such as Ti, V, Cr and Mn \cite{salemi2022first,jo2018gigantic,choi2023observation}. Experimentally, Cr has emerged as one of the most efficient orbital-current generators reported to date, while Pt has been identified as an excellent orbital-to-spin converter because of its strong SOC, making Pt- and Cr-based heterostructures particularly attractive platforms for spin-orbital torque generation~\cite{lee2021efficient,sala2022giant,gupta2025harnessing}.

To efficiently exert a torque on an adjacent ferromagnet, many previous experimental demonstrations have relied on engineering orbital-to-spin conversion either by employing ferromagnets with efficient orbital-to-spin conversion, such as Gd~\cite{sala2022giant}, or by inserting a thin Pt interfacial conversion layer between the orbital-current source and the ferromagnet~\cite{lee2021efficient,gupta2025harnessing}. While these approaches enable efficient orbital-to-spin conversion, they can introduce additional interface-dependent orbital-current transmission and spin-transparency parameters, making the resulting torque sensitive to interfacial orbital transport and conversion~\cite{sala2022giant,lyalin2024interface,zhang2015role}.
These considerations motivate the exploration of homogeneous materials in which orbital-current generation and orbital-to-spin conversion occur within a single layer before the transfer of angular momentum into the adjacent ferromagnet. Although orbital torques have recently enabled efficient magnetization control and current-induced switching~\cite{lee2021orbital,hayashi2023observation,lee2021efficient,gupta2025harnessing}, their ability to sustain coherent magnetization dynamics, especially auto-oscillations in nanoconstriction devices, remains unexplored.

Here, for the first time, we demonstrate spin-orbital Hall nano-oscillators (SOHNOs), as we term them, based on homogeneous PtCr/NiFe heterostructures. Our materials platform combines three complementary ingredients: \emph{i}) PtCr alloys that combine spin Hall currents from Pt with orbital Hall currents from Cr; \emph{ii}) the strong SOC of Pt, which facilitates orbital-to-spin conversion; and \emph{iii}) low-damping NiFe for sustaining current-driven auto-oscillations. Using dc-bias spin-torque ferromagnetic resonance (ST-FMR), electrical auto-oscillation measurements, structural characterization and first-principles calculations, we show that Cr alloying increases the effective torque efficiency from $\sim$0.14 in Pt/NiFe to $\sim$0.40 in Pt$_{0.38}$Cr$_{0.62}$/NiFe, while reducing the auto-oscillation threshold current density by nearly 60\%. These results establish PtCr as a promising platform for SOHNOs and expand the scope of orbitronics toward coherent magnetization dynamics.

\subsection*{Structural and magnetic characterization of PtCr/NiFe heterostructures}

\begin{figure*}[t]
\centering
\includegraphics[width=\textwidth]{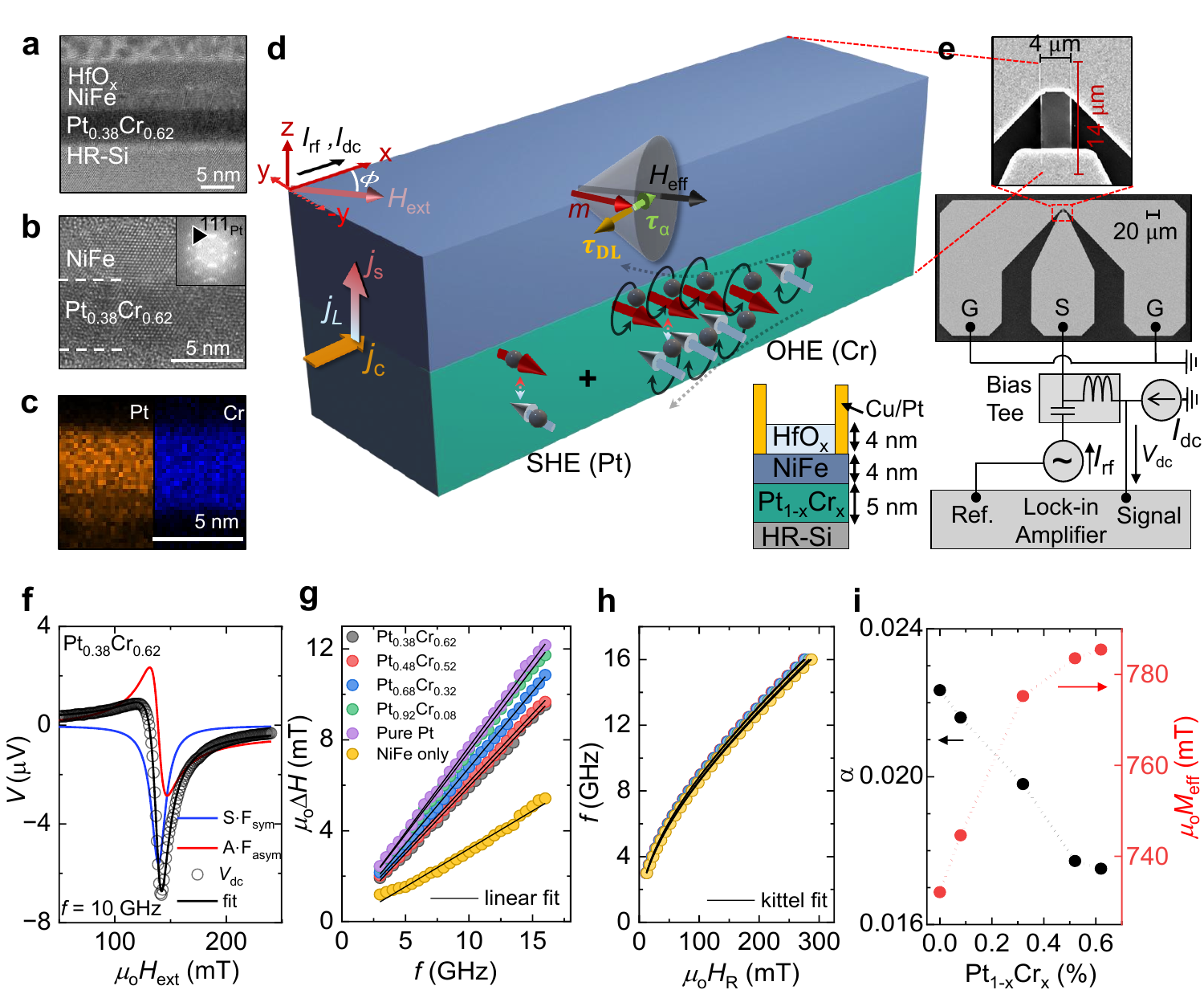}
\caption{
\textbf{Structure and magnetic properties of PtCr/NiFe heterostructures.}
\textbf{a}, Cross-sectional TEM image of the Pt$_{0.38}$Cr$_{0.62}$/NiFe heterostructure.
\textbf{b}, High-resolution TEM image of the Pt$_{0.38}$Cr$_{0.62}$/NiFe interface. Inset: corresponding FFT pattern with the fcc PtCr [111] diffraction spot marked, indicating an out-of-plane orientation close to the [111] direction.
\textbf{c}, STEM--EDXS elemental maps of Pt and Cr within the Pt$_{0.38}$Cr$_{0.62}$ layer.
\textbf{(d)}, Schematic illustration of the proposed spin--orbital angular-momentum transfer mechanism in PtCr/NiFe. A charge current density, $j_c$, generates a transverse spin current, $j_s$, in Pt through SHE and a transverse orbital current, $j_L$, in Cr through the OHE. Owing to the strong SOC of Pt, the orbital current is partially converted into a spin current, which exerts a $\tau_{\mathrm{DL}}$, on the magnetization, $m$, of the adjacent NiFe layer. The magnetization precesses about the effective magnetic field, $H_{\mathrm{eff}}$. The layer structure of the Pt$_{1-x}$Cr$_x$/NiFe heterostructure is also shown.
\textbf{e}, Scanning electron micrograph of the nanoconstriction device together with the dc-bias ST-FMR measurement configuration.
\textbf{f}, Representative ST-FMR spectrum measured at 10~GHz for Pt$_{0.38}$Cr$_{0.62}$/NiFe together with the fitted symmetric and antisymmetric Lorentzian components.
\textbf{g}, Linewidth as a function of microwave frequency for all compositions. Solid lines are linear fits.
\textbf{h}, Resonance frequency as a function of resonance field together with Kittel fits.
\textbf{i}, Extracted Gilbert damping parameter $\alpha$ (black) and effective magnetization $\mu_0M_{\mathrm{eff}}$ (red) as a function of Cr concentration.
}
\label{fig:stfmr}
\end{figure*}

We fabricated Pt$_{1-x}$Cr$_x$/NiFe heterostructures with varying Cr concentration to investigate spin-orbital angular momentum transport (see Supplementary Information 1 for details of thin-film deposition and sample preparation). We first characterize their structural and magnetic properties. The high structural quality is confirmed by the cross-sectional transmission electron microscopy (TEM) image of the Pt$_{0.38}$Cr$_{0.62}$/NiFe heterostructure shown in Fig.~\ref{fig:stfmr}a, revealing continuous multilayers with well-defined PtCr/NiFe interfaces. High-resolution TEM (HRTEM), together with the corresponding fast Fourier transform (FFT) pattern (Fig.~\ref{fig:stfmr}b), further confirm the crystalline nature of the PtCr layer. The FFT inset indicates an out-of-plane orientation close to the fcc PtCr [111] direction. The Scanning Transmission Electron Microscopy-Energy Dispersive X-ray Spectroscopy (STEM--EDXS) elemental mapping results (Fig.~\ref{fig:stfmr}c) verify the spatial distributions of Pt and Cr within the alloy layer without evidence of phase segregation on the measured length scale. Additional structural characterization results of other heterostructures, which were studied by TEM as well as grazing-incidence X-ray diffraction (GIXRD), are provided in Supplementary Information 2.

The proposed spin-orbital angular-momentum transfer mechanism is illustrated schematically in Fig.~\ref{fig:stfmr}d. A charge current flowing through the Pt component generates a transverse spin current via the SHE, whereas a charge current flowing through the Cr component generates a transverse orbital current via the OHE. Owing to the strong SOC of Pt, the orbital current is partially converted into a spin current before transferring angular momentum to the adjacent NiFe layer, providing an additional contribution to the damping-like torque, $\tau_{\mathrm{DL}}$. The layer structure of the Pt$_{1-x}$Cr$_x$/NiFe heterostructure is also illustrated in Fig.~\ref{fig:stfmr}d. Figure~\ref{fig:stfmr}e shows the scanning electron micrograph of the nanoconstriction device integrated with ground-signal-ground coplanar waveguides (GSG-CPW) together with the dc-bias ST-FMR measurement configuration. During the measurements, a microwave current $I_{\mathrm{rf}}$ excites magnetization precession, while the rectified mixing voltage $V_{\mathrm{dc}}$ is detected under a swept in-plane magnetic field applied at an angle of $\phi = 70^\circ$ with respect to the current direction.

A representative ST-FMR spectrum measured at 10~GHz is shown in Fig.~\ref{fig:stfmr}f. The measured voltage is fitted using \cite{liu2011spin}

\begin{equation}
V_{\mathrm{dc}}
=
S F_{\mathrm{sym}}(H_{\mathrm{ext}})
+
A F_{\mathrm{asym}}(H_{\mathrm{ext}}),
\end{equation}
where $F_{\mathrm{sym}}$ and $F_{\mathrm{asym}}$ denote the symmetric and antisymmetric Lorentzian components, respectively, while $S$ and $A$ are their corresponding amplitudes. The dominant symmetric contribution indicates a sizable $\tau_{\mathrm{DL}}$ arising from the combined spin Hall and orbital Hall channels. The linewidth ($\mu_0 \Delta H$) varies linearly with microwave frequency (Fig.~\ref{fig:stfmr}g), whereas the resonance frequency ($f$) follows the Kittel relation (Fig.~\ref{fig:stfmr}h), allowing the extraction of the Gilbert damping parameter $\alpha$ and effective magnetization $\mu_0M_{\mathrm{eff}}$ \cite{shashank2026bulk,shashank2025giant}. Experimental details of the ST-FMR analysis and extraction of magnetic parameters are provided in the Methods and Supplementary Information 3.

\subsection*{Enhanced current-induced torque in PtCr/NiFe heterostructures}

\begin{figure*}[t]
\centering
\includegraphics[width=\textwidth]{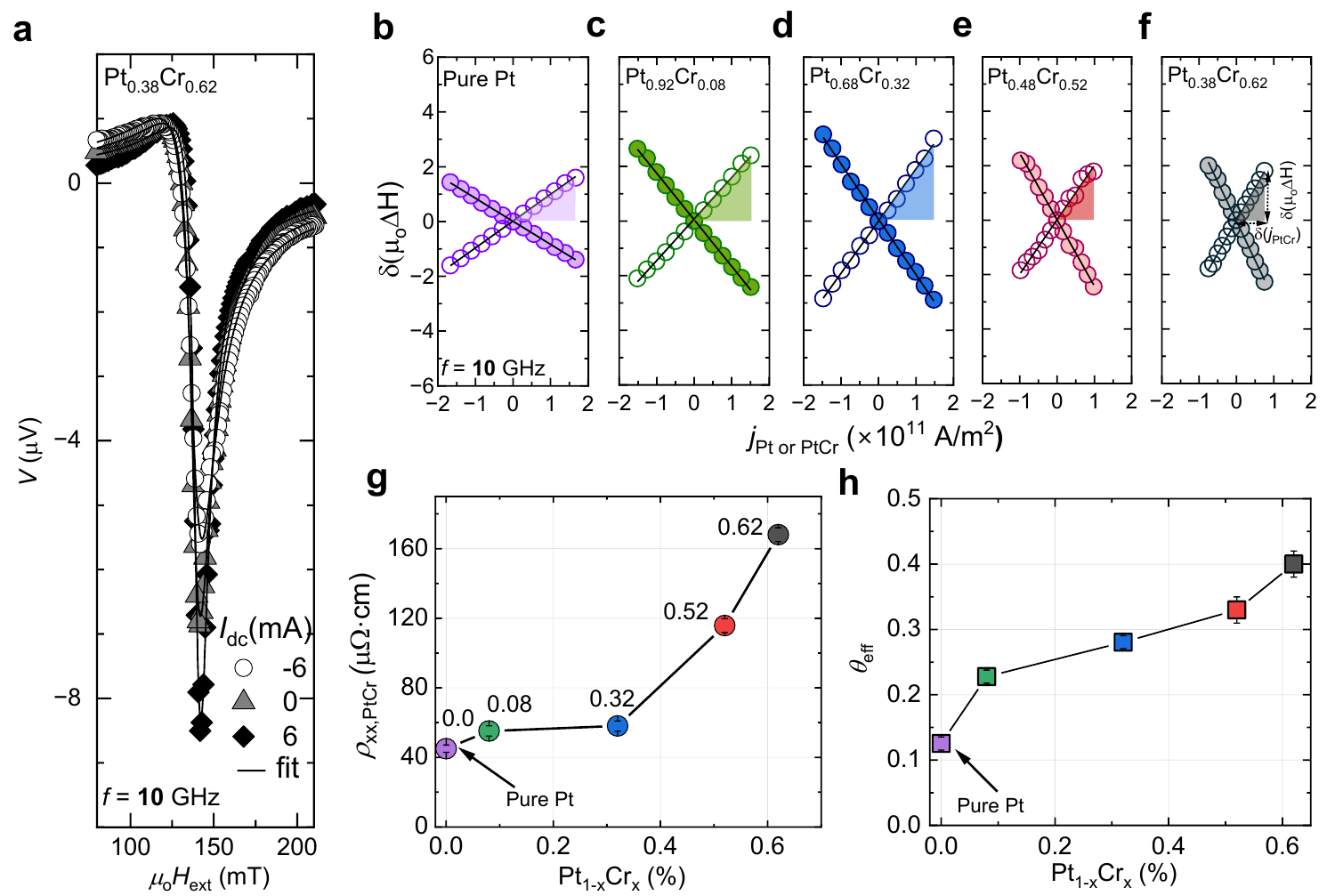}
\caption{
\textbf{Current-induced torque generation in PtCr/NiFe heterostructures.}
\textbf{a}, Representative ST-FMR spectra measured under different dc bias currents for Pt$_{0.38}$Cr$_{0.62}$/NiFe at $f$ = 10 GHz. The black solid lines represent fits to Eq.~(1). 
\textbf{b--f}, Linewidth modulation, $\delta(\mu_0\Delta H)$, as a function of current density for Pt/NiFe and Pt$_{1-x}$Cr$_x$/NiFe measured at $f$ = 10 GHz and at $\phi = 70^\circ$. Solid lines are linear fits. Open and filled circles represent positive and negative magnetic-field polarities, respectively.
\textbf{g}, Electrical resistivity of Pt$_{1-x}$Cr$_x$ as a function of Cr concentration, determined from four-probe measurements on blanket films and two-probe measurements on patterned devices. Error bars represent one standard deviation across repeated measurements.
\textbf{h}, Effective torque efficiency, $\theta_{\mathrm{eff}}$, extracted from dc-bias ST-FMR. Error bars represent the propagated uncertainty from the ST-FMR fitting and current-density determination using Eqs. (1), (2) and (3).
}
\label{fig:dcstfmr}
\end{figure*}

To quantify the current-induced torque, we performed dc-bias ST-FMR, in which a dc current is applied simultaneously with the microwave excitation. The dc current modifies the effective magnetic damping through the $\tau_{\mathrm{DL}}$, resulting in a current-dependent modulation of the resonance linewidth \cite{liu2011spin}. Representative ST-FMR spectra measured under different dc bias currents for Pt$_{0.38}$Cr$_{0.62}$/NiFe are shown in Fig.~\ref{fig:dcstfmr}a. As the dc current is varied from $-6$ to $+6$ mA, a systematic modulation of the resonance linewidth is clearly observed. The change in linewidth is quantified as

\begin{equation}
\delta(\mu_0\Delta H)
=
\mu_0\Delta H(I_{\mathrm{dc}})
-
\mu_0\Delta H(I_{\mathrm{dc}}=0),
\label{eq:linewidth}
\end{equation}
which varies linearly with $I_{\mathrm{dc}}$. The effective torque efficiency, $\theta_{\mathrm{eff}}$, was determined from the absolute slope of the linewidth modulation,
$\vert \Delta[\delta(\mu_0\Delta H)] \vert / \vert \Delta j_{\mathrm{dc,PtorPtCr}} \vert$, according to \cite{liu2011spin}

\begin{equation}
\theta_{\mathrm{eff}}
=
\frac{2e}{\hbar}
\frac{(H_R+M_{\mathrm{eff}}/2)\mu_0M_{\mathrm{s}}t}{\sin\phi}
\frac{\gamma}{2\pi f}
\frac{\vert \Delta[\delta(\mu_0\Delta H)] \vert}
{\vert \Delta j_{\mathrm{dc,Pt or PtCr}} \vert},
\label{eq:thetaeff}
\end{equation}
where $H_R$ is the resonance field, $M_{\mathrm{eff}}$ is the effective magnetization extracted from the Kittel analysis \cite{shashank2026bulk}, and $M_{\mathrm{s}}$ and $t$ are the saturation magnetization and thickness of the NiFe layer, respectively. Here, $\phi = 70^\circ$ is the angle between the applied dc current and the external magnetic field. Furthermore, $\gamma$ is the gyromagnetic ratio, $f$ is the microwave frequency, and $j_{\mathrm{dc,Pt or PtCr}}$ is the dc current density flowing in the Pt or PtCr layer, respectively estimated using the measured resistivity and a parallel-resistor current-shunting model. The extracted linewidth modulation, $\delta(\mu_0\Delta H)$, exhibits an approximately linear dependence on the dc current density for all Pt$_{1-x}$Cr$_x$/NiFe heterostructures, as shown in Fig.~\ref{fig:dcstfmr}b--f. The systematic increase in slope with increasing Cr concentration indicates progressively enhanced current-induced damping-like torque.

The electrical resistivity of Pt$_{1-x}$Cr$_x$ increases monotonically with Cr concentration, as shown in Fig.~\ref{fig:dcstfmr}g. The extracted effective torque efficiencies, $\theta_{\mathrm{eff}}$, are summarized in Fig.~\ref{fig:dcstfmr}h. $\theta_{\mathrm{eff}}$ increases systematically from 0.14 for Pt/NiFe to approximately 0.40 for Pt$_{0.38}$Cr$_{0.62}$/NiFe, corresponding to a nearly threefold enhancement despite substantial Pt dilution. Transparency-corrected spin--orbital pumping voltage signals, obtained using the standard spin-pumping formalism, exhibit the same composition-dependent trend (see Supplementary Information 4 for details of spin-orbital pumping measurements and analysis), confirming that the enhanced torque is robust against interfacial spin-transmission corrections and is consistent with recent experimental demonstrations of reciprocal spin-orbital pumping~\cite{hayashi2024observation,keller2025identification}. Although the spin Hall response has been reported to differ between sputtered and evaporated Pt films prepared under different deposition conditions~\cite{Sagasta2016}, the present PtCr layers remain crystalline, as confirmed by GIXRD and cross-sectional HRTEM (see Supplementary Information 2). Therefore, the monotonic enhancement of $\theta_{\mathrm{eff}}$ cannot be readily attributed to amorphization or disorder-driven extrinsic scattering alone~\cite{wang2022giant,shashank2025giant}. Moreover, the enhancement in $\theta_{\mathrm{eff}}$ occurs simultaneously with a reduction in the Gilbert damping. While the measured damping contains multiple relaxation contributions, its systematic reduction with increasing Cr concentration is consistent with a reduced spin-pumping contribution arising from Pt dilution, as Pt is an efficient spin sink that enhances Gilbert damping through spin pumping~\cite{tserkovnyak2002enhanced,shashank2023disentanglement}. However, this mechanism alone cannot account for the concurrent increase in the effective torque efficiency. Instead, these observations are consistent with an additional orbital-mediated angular-momentum transfer pathway beyond the conventional SHE, as illustrated schematically in Fig.~\ref{fig:stfmr}d. Such a combination of enhanced torque efficiency and reduced magnetic damping is expected to lower the threshold for current-driven auto-oscillations, which we investigate in the following section.

\subsection*{Spin-Orbital Hall nano-oscillators in PtCr/NiFe heterostructures}

Having established the enhanced torque efficiency and reduced magnetic damping in PtCr/NiFe, we next investigate whether these improvements translate into enhanced nonlinear magnetization dynamics. To this end, we implement the PtCr/NiFe heterostructures in the well-established nanoconstriction SHNO geometry, which has enabled coherent auto-oscillations, propagating spin waves, mutual synchronization in chains and arrays, and large-scale oscillator networks~\cite{demidov2014nanoconstriction,houshang2016spinwave,Awad2017,fulara2019spin,kumar2025spin,behera2026nanosecond}. Using this platform, we realize SOHNOs based on PtCr/NiFe heterostructures.

\begin{figure*}[t]
\centering
\includegraphics[width=\textwidth]{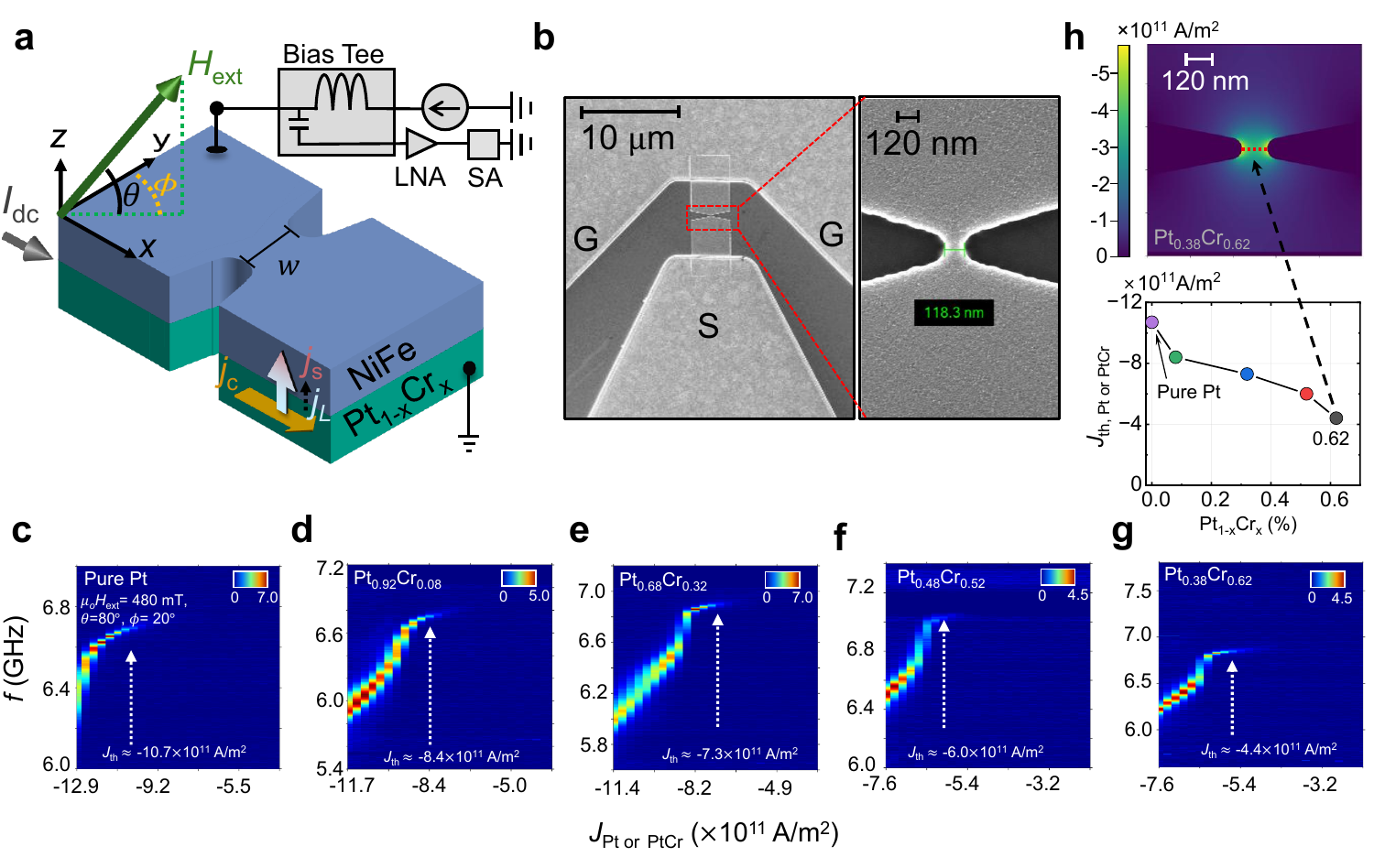}
\caption{
\textbf{Spin-Orbital Hall nano-oscillators based on PtCr/NiFe heterostructures.}
\textbf{a}, Schematic of the nanoconstriction-based nano-oscillator measurement geometry.
\textbf{b}, Scanning electron micrographs of the nanoconstriction device. Insets show the enlarged constriction region with a nominal width of 120~nm.
\textbf{c--g}, Power spectral density (PSD) as a function of current density measured under an applied magnetic field of $\mu_0H_{\mathrm{ext}}=480$~mT ($\theta=80^\circ$, $\phi=20^\circ$) for Pt/NiFe, Pt$_{0.92}$Cr$_{0.08}$/NiFe, Pt$_{0.68}$Cr$_{0.32}$/NiFe, Pt$_{0.48}$Cr$_{0.52}$/NiFe, and Pt$_{0.38}$Cr$_{0.62}$/NiFe, respectively. White arrows indicate the threshold current density for the onset of coherent auto-oscillations, while the color bars indicate the peak power (dB above noise floor).
\textbf{h}, Top panel shows the simulated current-density distribution in the 120-nm-wide nanoconstriction. The red dashed line indicates the center line of the nanoconstriction over which the simulated current density was averaged to determine the threshold current density discussed in the text. Bottom panel shows the experimentally extracted threshold current density, $J_{\mathrm{th,\;\text{Pt or PtCr}}}$, as a function of Cr concentration.
}
\label{fig:shno}
\end{figure*}

Figures~\ref{fig:shno}a,b show the measurement geometry and scanning electron micrographs of the 120-nm-wide nanoconstriction devices. A direct current was applied through a bias tee, while the generated microwave signals were extracted from the high-frequency port, amplified using a low-noise amplifier, and analyzed with a spectrum analyzer. Auto-oscillation measurements were performed under an external magnetic field of $\mu_0H_{\mathrm{ext}}=480$~mT applied at an out-of-plane angle of $\theta=80^\circ$ and an in-plane angle of $\phi=20^\circ$, where $\phi$ is defined with respect to the device axis. In nanoconstriction SHNOs, the geometrically enhanced local current density enables the $\tau_{\mathrm{DL}}$ to balance the intrinsic damping torque, $\tau_{\alpha}$, thereby fulfilling the threshold condition for sustained auto-oscillations~\cite{demidov2014nanoconstriction,Awad2017}. Interestingly, as $\alpha$ decreases systematically with increasing Cr concentration (Fig.~\ref{fig:stfmr}i), strong microwave emission is preserved across the entire composition series, indicating that the enhanced $\tau_{\mathrm{DL}}$ complements the reduced $\alpha$, which is favourable for auto-oscillation.

Figures~\ref{fig:shno}c--g show the evolution of the power spectral density as a function of current density for Pt/NiFe and Pt$_{1-x}$Cr$_x$/NiFe heterostructures. All devices exhibit coherent microwave emission above a well-defined threshold current density. With increasing Cr concentration, the onset of auto-oscillations shifts systematically towards lower current densities while maintaining stable single-mode emission.

The extracted threshold current density, $J_{\mathrm{th,\;Pt\,or\,PtCr}}$ decreases continuously from approximately $1.07\times10^{12}\,\mathrm{A\,m^{-2}}$ in Pt/NiFe to approximately $4.4\times10^{11}\,\mathrm{A\,m^{-2}}$ in Pt$_{0.38}$Cr$_{0.62}$/NiFe (Fig.~\ref{fig:shno}h, lower panel), corresponding to a reduction of nearly 60\%, where the averaging was done over the red central line of the nanoconstriction (Fig.~\ref{fig:shno}h, upper panel), perpendicular to the current direction (see Supplementary Information 5 for details on current density simulations). This substantial reduction follows the combined trend of enhanced $\theta_{\mathrm{eff}}$ and reduced $\alpha$  established in Figs.~\ref{fig:dcstfmr}h and~\ref{fig:stfmr}i, respectively. The simultaneous enhancement of $\theta_{\mathrm{eff}}$ and reduction of magnetic damping therefore translate directly into lower threshold current densities for sustained auto-oscillations. Together with the reciprocal orbital-pumping measurements presented above, these results establish PtCr/NiFe heterostructures as an efficient platform for orbitronic nano-oscillators.

\subsection*{First-principles evidence for a mixed spin--orbital torque mechanism}

\begin{figure*}[t]
\centering
\includegraphics[width=\textwidth]{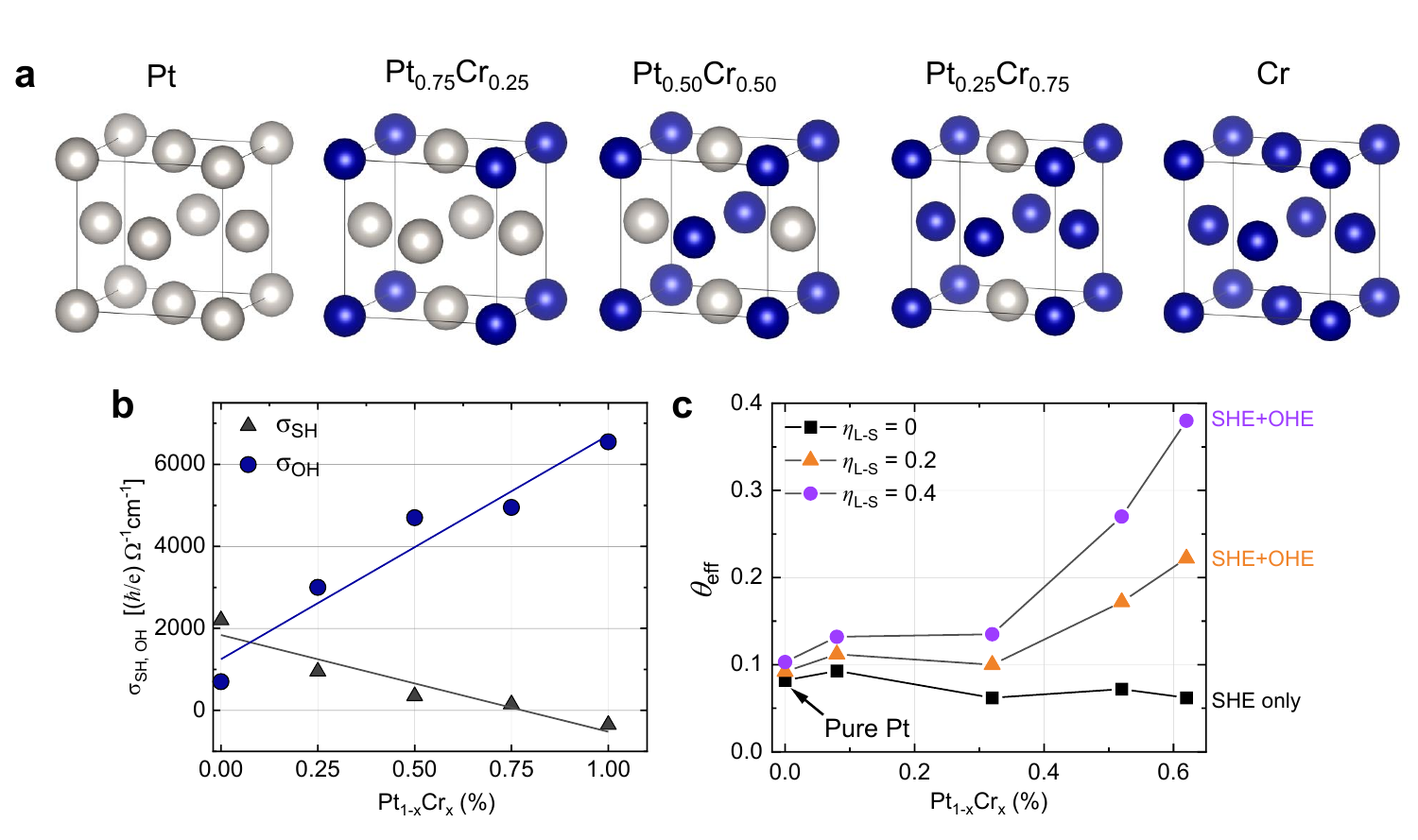}
\caption{
\textbf{First-principles evidence for a mixed spin--orbital torque mechanism in PtCr.}
\textbf{a}, Fully relaxed Pt$_{1-x}$Cr$_x$ supercell structures used for first-principles calculations.
\textbf{b}, Calculated intrinsic spin Hall conductivity, $\sigma_{\mathrm{SH}}$, and intrinsic orbital Hall conductivity, $\sigma_{\mathrm{OH}}$, as a function of Cr concentration. Solid lines are linear fits used to interpolate the conductivities at the experimental compositions.
\textbf{c}, Estimated effective torque efficiency calculated using Eq.~(\ref{eq:dft_theta}), where $\rho_{\mathrm{PtCr}}(x)$ is the experimentally measured PtCr resistivity and $\eta_{\mathrm{L-S}}$ is an effective orbital-to-spin conversion parameter. The SHE-only limit corresponds to $\eta_{\mathrm{L-S}}=0$, while finite $\eta_{\mathrm{L-S}}$ incorporates orbital-to-spin conversion of the orbital Hall current. The experimentally observed increase in $\theta_{\mathrm{eff}}$ is reproduced only when the orbital Hall contribution is included.
}
\label{fig:dft}
\end{figure*}

Next, the origin of the enhanced $\theta_{\mathrm{eff}}$ is investigated. We performed first-principles calculations of the intrinsic spin Hall conductivity, $\sigma_{\mathrm{SH}}(x)$, and orbital Hall conductivity, $\sigma_{\mathrm{OH}}(x)$, in Pt$_{1-x}$Cr$_x$ alloys. Since the PtCr layers remain crystalline, as established by GIXRD and HRTEM, the alloys were modeled using fcc-based supercells (Fig.~\ref{fig:dft}a). Extrinsic contributions from disorder scattering were neglected because the spin and orbital Hall conductivities of transition metals are dominated by intrinsic contributions at room temperature~\cite{mankovsky2024spin}. Moreover, the extrinsic SHE in Pt alloys is predicted to be negligible except in the dilute regime ($x < 0.05$)~\cite{lowitzer2011extrinsic}. Computational details are provided in Methods. As shown in Fig.~\ref{fig:dft}b, the calculated $\sigma_{\mathrm{SH}}$ of pure Pt is of the order of $10^3~(\hbar/e)\,\Omega^{-1}\,\mathrm{cm}^{-1}$, consistent with previous theoretical and experimental reports~\cite{Sagasta2016,salemi2022first}. With increasing Cr concentration, $\sigma_{\mathrm{SH}}$ decreases, whereas $\sigma_{\mathrm{OH}}$ increases monotonically. Such monotonic composition dependences allow $\sigma_{\mathrm{SH}}(x)$ and $\sigma_{\mathrm{OH}}(x)$ to be reasonably estimated for arbitrary $x$ using linear fits.

The effective torque efficiency, including both spin and orbital Hall contributions, is then estimated using~\cite{go2020orbital,lee2021orbital,Bai2025}

\begin{equation}
\theta_{\mathrm{eff}}(x)
=
\rho_{\mathrm{PtCr}}(x)
\left[
\sigma_{\mathrm{SH}}(x)
+
\eta_{\mathrm{L-S}}
\sigma_{\mathrm{OH}}(x)
\right],
\label{eq:dft_theta}
\end{equation}

where $\rho_{\mathrm{PtCr}}(x)$ is the experimentally measured resistivity of Pt$_{1-x}$Cr$_x$, and $\eta_{\mathrm{L-S}}$ is a phenomenological parameter describing the conversion efficiency of orbital currents into spin currents. This orbital-to-spin conversion can be substantial due to the strong SOC of Pt. Because it saturates within approximately 1~nm of Pt~\cite{ding2020harnessing,bose2023detection}, $\eta_{\mathrm{L-S}}$ is treated as a constant independent of $x$. The limit $\eta_{\mathrm{L-S}}=0$ corresponds to a pure spin Hall scenario.

As shown in Fig.~\ref{fig:dft}c, the SHE-only model predicts little variation in $\theta_{\mathrm{eff}}$ with Cr concentration and therefore fails to reproduce the experimentally observed enhancement. This is because the increase in $\rho_{\mathrm{PtCr}}(x)$ with $x$ (Fig.~\ref{fig:dcstfmr}g) is offset by the decrease in $\sigma_{\mathrm{SH}}(x)$ (Fig.~\ref{fig:dft}b), indicating that an additional contribution beyond the SHE is required to account for the substantial enhancement of $\theta_{\mathrm{eff}}$. In contrast, including a finite orbital-to-spin conversion naturally yields a monotonic increase in $\theta_{\mathrm{eff}}$, in excellent agreement with the dc-bias ST-FMR measurements. These results support a mixed spin--orbital transport mechanism in which the increasingly large orbital Hall current generated by Cr is efficiently converted into a spin current via the strong SOC of Pt.

We note that enhanced $\theta_{\mathrm{eff}}$ in PtCr-based heterostructures has also been reported previously. Hu \textit{et al.}~\cite{hu2022toward} discussed several possible microscopic origins for the enhanced torque efficiency, including orbital-to-spin conversion associated with the OHE together with resistivity-enhanced SHE, without uniquely identifying the dominant angular-momentum transport mechanism, whereas Liu \textit{et al.}~\cite{liu2022strong} attributed the torque observed in their Cr$_{1-x}$Pt$_x$/Co heterostructures to an intrinsic SHE, demonstrating that their experimental results could be quantitatively reproduced within this framework. In contrast, the substantially larger enhancement observed here cannot be readily accounted for within an SHE-only picture. More recently, elegant heterostructure engineering approaches have demonstrated efficient orbital-torque generation by enhancing orbital-to-spin conversion through high-SOC ferromagnets and Pt interfacial conversion layers in Cr-based systems~\cite{lee2021efficient,gupta2025harnessing}, significantly advancing the experimental realization of orbitronic devices. Building on these developments, the present PtCr/NiFe bilayer integrates orbital-current generation and orbital-to-spin conversion within a single crystalline alloy before angular momentum is transferred directly to the adjacent NiFe layer, thereby eliminating the need for a dedicated conversion layer and reducing the number of interfaces involved in angular-momentum transport. Furthermore, our 4-nm NiFe control exhibits a negligible self-induced $\tau_{\mathrm{DL}}$ contribution (Supplementary Information 6)~\cite{seki2021spin}. Together with the reciprocal spin--orbital pumping measurements and first-principles calculations, these observations consistently support a mixed spin--orbital transport mechanism beyond a purely SHE-based interpretation, establishing homogeneous alloy engineering as a complementary route toward efficient orbital-torque devices and coherent spin-orbital Hall nano-oscillators.

\section*{Conclusion}\label{sec7}

In summary, we demonstrate, for the first time, spin--orbital Hall nano-oscillators (SOHNOs) based on Pt$_{1-x}$Cr$_x$/NiFe heterostructures. Cr alloying simultaneously enhances the effective torque efficiency from 0.14 to 0.40, strengthens the reciprocal spin--orbital pumping response, reduces the effective Gilbert damping by approximately 23\%, and lowers the auto-oscillation threshold current density by nearly 60\%, while only modestly increasing $\mu_0M_{\mathrm{eff}}$ by about 7\%. Combined with first-principles calculations, our results show that the experimentally observed torque enhancement cannot be reproduced by the intrinsic spin Hall contribution alone, but emerges naturally when the orbital Hall contribution is incorporated. These findings provide compelling evidence for a mixed spin--orbital transport mechanism in crystalline PtCr, where Pt supplies efficient spin Hall conversion and strong spin--orbit coupling, while Cr contributes an increasingly large orbital Hall current. Together, these results establish crystalline PtCr as an efficient platform for SOHNOs and highlight alloy engineering as a general strategy for exploiting coupled spin and orbital transport in low-power orbitronic and spintronic devices.

Looking ahead, important challenges \cite{fukami2025challenges} include the quantitative disentanglement of spin and orbital current contributions, accurate determination of orbital transport parameters, and a microscopic understanding of orbital transmission across interfaces. Addressing these questions and extending the present strategy to other transition-metal alloys and low-damping magnetic heterostructures may enable a new generation of energy-efficient orbitronic devices, ranging from SOHNOs and coherent spin-wave emitters to scalable oscillator networks and unconventional computing architectures~\cite{Awad2017,fulara2019spin,behera2026nanosecond,jo2024spintronics}.

\section*{Methods}

\subsection*{Thin-film preparation}

Pt$_{1-x}$Cr$_x$(5 nm)/NiFe(4 nm)/HfO$_x$(4 nm) heterostructures were deposited on high-resistance Si (HR-Si) substrates ($\rho > 10{,}000~\Omega\cdot\mathrm{cm}$) by DC/RF magnetron sputtering (AJA Orion 8) under a base pressure below $5\times10^{-8}$ mbar. $\mathrm{Pt}_{1-x}\mathrm{Cr}_{x}$ alloys with controlled Cr content were synthesized by DC co-sputtering of Pt and Cr targets at an Ar pressure of 3~mTorr. The alloy composition was tuned by varying the Pt sputtering power while adjusting the Cr power accordingly. The thickness of PtCr was kept constant at 5 nm. Pure Pt, NiFe, and HfO$_x$ layers were deposited under identical conditions. The Pt$_{1-x}$Cr$_x$ composition was determined from the calibrated deposition rates, densities, and molar masses of Pt and Cr, yielding their respective molar deposition rates. Details of the alloy calibration procedure, sputtering-rate measurements, and composition determination are provided in Supplementary Information 1.

\noindent

\subsection*{Theory of spin--orbital Hall transport}

First-principles calculations were performed within density functional theory using the Quantum ESPRESSO package~\cite{giannozzi2009quantum}. The Perdew--Burke--Ernzerhof functional~\cite{perdew1996generalized} within the generalized gradient approximation was used for the exchange-correlation functional. The ionic potentials were described by fully relativistic projector augmented-wave pseudopotentials~\cite{kresse1999from} from the pslibrary database~\cite{corso2014pseudopotentials}. Spin-orbit coupling was included self-consistently. The kinetic energy cutoff for the wave functions and the charge density cutoff were set to 80 Ry and 640 Ry, respectively. The Pt$_{1-x}$Cr$_x$ alloys were modeled using fcc-based supercells, as shown in Fig. 4a. For each composition, the lattice parameter and atomic positions were fully relaxed until the forces on all atoms were smaller than $10^{-4} \ \mathrm{Ry}/a_0$. For self-consistent calculations, a $12 \times 12 \times 12$ $k$-point mesh was used. 

Maximally localized Wannier functions~\cite{souza2001maximally} were constructed using the Wannier90 code~\cite{pizzi2020wannier90}. The initial orbital projections were chosen as $s$, $p$, and $d$ orbitals for each Pt and Cr atom. The frozen energy window for disentanglement extended from the bottom of the valence bands to 5 eV above the Fermi level. For the linear-response calculations, the Hamiltonian ($\hat{H}$), spin ($\hat{\mathbf{S}}$), orbital angular momentum ($\hat{\mathbf{L}}$), and position ($\hat{\mathbf{r}}$) operators were interpolated on a dense $100 \times 100 \times 100$ $k$-point mesh, where the orbital angular momentum operator was defined in an atomic-like real-harmonic basis. 

The intrinsic spin Hall conductivity ($\sigma_\mathrm{SH}$) and orbital Hall conductivity ($\sigma_\mathrm{OH}$) were calculated within linear-response theory using the Kubo formula~\cite{go2018intrinsic, jo2018gigantic, salemi2022first}:
\begin{equation}
	\sigma_\mathrm{SH(OH)} = 
	-e\hbar \int \frac{d^3\mathbf{k}}{(2\pi)^3} \sum_{n \neq m} 
	(f_{n\mathbf{k}} - f_{m\mathbf{k}}) 
	\mathrm{Im}
	\left[
	\frac
	{\langle \psi_{n\mathbf{k}} \vert \hat{J}_z^{X_y}  \vert \psi_ {m\mathbf{k}}\rangle 
		\langle \psi_{m\mathbf{k}} \vert \hat{v}_x \vert \psi_{n\mathbf{k}} \rangle 
	}
	{(\varepsilon_{n\mathbf{k}} - \varepsilon_{m\mathbf{k}})( \varepsilon_{n\mathbf{k}} - \varepsilon_{m\mathbf{k}} + i\Gamma )} 
	\right], 
\end{equation}
where $\vert \psi_{n\mathbf{k}} \rangle$ is the eigenstate of band $n$, $\varepsilon_{n\mathbf{k}}$ is the corresponding energy eigenvalue, $f_{n\mathbf{k}} $ is the Fermi--Dirac distribution function at room temperature. The spin and orbital current operators are defined as $\hat{J}_z^{X_y} = \frac{1}{2} (\hat{v}_z \hat{X}_y + \hat{X}_y \hat{v}_z)$, where $X_y = S_y$ for $\sigma_\mathrm{SH}$ and $X_y = L_y$ for $\sigma_\mathrm{OH}$. The velocity operator is given by  $\hat{\mathbf{v}} = \frac{1}{i\hbar} [\hat{\mathbf{r}}, \hat{H}]$. The lifetime broadening energy $\Gamma$ was set to 0.1~eV.

\subsection*{Grazing-incidence X-ray diffraction}

The crystal structure of the Pt$_{1-x}$Cr$_x$/NiFe heterostructures was characterized by grazing-incidence X-ray diffraction (GIXRD) using a Mat:Nordic SAXSLAB instrument equipped with a Rigaku 003 microfocusing Cu K$\alpha$ X-ray source, a two-bounce monochromator, and Dectris Pilatus3 300K R and 100K detectors. The beam path was evacuated to minimize air scattering, and a grazing-incidence angle of $1^\circ$ was employed to maximize the diffraction signal from the thin films. Two-dimensional diffraction patterns were processed using SAXSGUI. Peak positions, lattice parameters, and full-width at half maximum (FWHM) values were extracted using TOPAS v6 (Bruker AXS) by fitting the diffraction peaks with a pseudo-Voigt (PVII) profile based on the fcc Pt structural model (space group $Fm\bar{3}m$). Instrumental broadening was determined using a LaB$_6$ standard and deconvoluted from the measured peak widths. The exposure time for each measurement was 3 h.

\subsection*{Cross-sectional TEM and STEM-EDXS}

The microstructure and chemical composition of the heterostructures were investigated by cross-sectional transmission electron microscopy (TEM), high-resolution TEM (HRTEM), and scanning transmission electron microscopy energy-dispersive X-ray spectroscopy (STEM-EDXS) using a JEOL monochromated ARM200F microscope operated at 200 kV. The microscope is equipped with a Schottky field-emission gun, double-Wien monochromator, probe and image aberration correctors, a Gatan GIF Continuum energy filter, a Gatan OneView camera, and dual silicon-drift detectors for EDXS analysis. Cross-sectional TEM lamellae were prepared using a FEI Versa 3D focused ion beam/scanning electron microscope (FIB-SEM). After depositing a protective Pt/C layer by electron-beam and Ga-ion-beam deposition, the lamellae were lifted out and progressively thinned using a 30 kV Ga ion beam with decreasing beam current from 1 nA to 100 pA. Final polishing was performed at 5 and 2 kV to minimize ion-beam-induced damage.

\subsection*{Device fabrication}

ST-FMR microstrips ($14\times4~\mu\mathrm{m}^2$), spin-orbital pumping microbars ($400\times2~\mu\mathrm{m}^2$), and 120-nm-wide nanoconstriction spin-orbital Hall nano-oscillators (SOHNOs) were fabricated on the same chip using electron-beam lithography (Raith EBPG 5200) followed by Ar-ion milling. GSG-CPW were subsequently fabricated by mask-less ultraviolet lithography (Heidelberg Instruments MLA 150) and lift-off, followed by Cu(800 nm)/Pt(20 nm) deposition by DC magnetron sputtering. Details of the fabrication procedure are provided in Ref.~\cite{shashank2026bulk}.

\subsection*{ST-FMR, spin--orbital pumping and auto-oscillation measurements}

ST-FMR measurements were performed using a PhaseFMR-40ST (NanOsc Instruments AB). Spin--orbital pumping measurements were performed on $400 \times 2~\mu\mathrm{m}^2$ microbars. Microwave currents were generated using a Rohde \& Schwarz SMB100A microwave source, while the rectified voltage, $V_{\mathrm{ISHE+IOHE}}$, was detected using a Stanford Research Systems SR830 lock-in amplifier under a rotatable in-plane magnetic field. The measurements employed an out-of-plane Oersted-field excitation geometry, as described in Ref. \cite{shashank2021enhanced}. Details of the ST-FMR fitting procedure and spin--orbital pumping analysis are provided in the Supplementary Information 3 and 4.

Auto-oscillation measurements were carried out on 120-nm-wide nanoconstriction SOHNOs by applying a dc current through a bias tee using a Keithley 2400 source meter. The emitted microwave signal was extracted from the high-frequency port, amplified using a Low Noise Factory (LNF) LNR4-14B low-noise amplifier, and analyzed using a Rohde \& Schwarz FSV spectrum analyzer with a resolution bandwidth of 1~MHz. All electrical measurements were performed at room temperature. More details are provided in Ref. \cite{shashank2026bulk,kumar2023robust,behera2026nanosecond,zahedinejad2020two}
\section*{Declarations}

\section*{Acknowledgements}

This work was supported by the Knut and Alice Wallenberg Foundation through Grant Nos. 2022.0079 (P.M.O. and J.~Å.) and 2023.0336 (P.M.O.). U.S. acknowledges support from the Marie Skłodowska-Curie Postdoctoral Fellowship ``GOHE'' (Grant Agreement No.~101152484). P.M.O. acknowledges support from the EIC Pathfinder OPEN Grant No. 101129641 (OBELIX). S.G. acknowledges support from the Marie Sk{\l}odowska-Curie Postdoctoral Fellowship ``MANGA'' (Grant Agreement No.~101152006). Financial support from the Swedish Research Council (VR) and the Swedish Foundation for Strategic Research (SSF) for access to ARTEMI, the Swedish National Infrastructure in Advanced Electron Microscopy (Grant Nos. 2021-00171 and RIF21-0026), is gratefully acknowledged. This work was performed in part at the Chalmers Materials Analysis Laboratory (CMAL). The calculations were enabled by resources provided by the National Academic Infrastructure for Supercomputing in Sweden (NAISS) at NSC Linköping, partially funded by the Swedish Research Council through Grant Agreement No. 2022-06725.

\bmhead{Conflict of interest}

The authors declare no competing interests.

\bmhead{Ethics approval}
Not applicable

\bmhead{Consent to participate}
Not applicable

\bmhead{Consent for publication}
Not applicable

\bmhead{Availability of data and materials}
The data is available upon reasonable request from corresponding authors.

\bmhead{Code availability}
Not applicable

\bmhead{Authors' contributions}

U.S., A.K., D.J., P.M.O., and J.\AA{}. conceived and planned the study. T.N.A.N. and U.S. optimized the PtCr thin-film growth, and T.N.A.N. prepared the thin films. U.S. fabricated the devices in discussion with A.K. J.-G.C. captured the SEM images. S.G. designed the spin-orbital pumping devices in discussion with U.S. U.S. performed all the spin-torque ferromagnetic resonance, spin-orbital pumping, and electrical auto-oscillation measurements and analyzed the data in  discussion with A.K., J.-G.C., A.A.A., and J.\AA{}. M.S. performed the GIXRD measurements. R.K. carried out the current-density simulations. L.Z. and A.B.Y. performed the TEM measurements with input from E.O. D.J. and P.M.O. carried out the first-principles calculations and developed the theoretical interpretation of the spin-orbital torque mechanism in discussion with J.\AA{}., A.K. and U.S. J.\AA{}. and P.M.O. managed the project. All authors contributed to the analysis and interpretation of the data and co-wrote the manuscript.


\bibliography{references}


\end{document}